\documentclass[amsmath,amssymb,aps,prb,twocolumn]{revtex4-2}
\usepackage[utf8]{inputenc}
\usepackage{graphicx,hyperref}
\usepackage{siunitx}

\begin{document}
\title{Revealing an anisotropic electronic scattering rate in the ``non-metallic" metal FeCrAs using the Hall effect}
\author{B. Lau}
\author{W. Wu}
\author{S. R. Julian}
\affiliation{Department of Physics, University of Toronto, 60 St. George Street, Toronto, Ontario, Canada M5S 1A7}
\date{\today}
\begin{abstract}
 We probe the relationship between magnetism and ``non-metallic" transport in the magnetically frustrated metal FeCrAs using the low-field Hall effect. We find that the low-field Hall coefficient is dependent on crystal orientation, and exhibits two temperature-induced sign reversals when the current is within the $ab$-plane. We suggest that these observations are due to scattering from magnetic fluctuations that become momentum-dependent near the onset of long-range magnetic order. It seems, however, that this additional spin-fluctuation scattering rides on top of the non-metallic resistivity, and Hund's metal interorbital correlations remain a possible source of the non-metallic behaviour. 
\end{abstract}

\maketitle
\newpage 
\section{Introduction}

In metals, 
  anomalous temperature dependence of transport properties, 
  from the Kondo resistivity minimum 
  to strange metal behaviour of cuprates, 
  is often a tell-tale sign of 
  strong correlations.
A conventional hallmark of metallic conductivity 
  is
  that the electrical resistivity falls with 
  decreasing temperature.
This behaviour has two causes.
The first is freezing out of 
  thermally excited modes, 
  such as phonons, 
  that electrons scatter from.
The second is a 
  decreasing phase space for 
  electrons to scatter into, 
  as the Fermi-Dirac distribution 
  contracts onto the Fermi surface.
When this conventional behaviour fails, 
  it means that a gap is opening on the 
  Fermi surface and/or that the scattering cross section from 
  impurities or fluctuating modes is 
  increasing with decreasing temperature.

    FeCrAs is a system with unique properties - its resistivity behaves in an unconventional ``non-metallic" manner over a huge 
    temperature range, and its low-temperature dependence is inconsistent with the usual Fermi liquid description~\cite{wu2009novel}. In particular, the resistivity of FeCrAs rises gradually with decreasing temperature from the highest ($>$~800 K) to the lowest ($<$~100 mK) temperatures measured, and approaches 0 K with a sub-linear power law temperature dependence. Despite these curious properties, FeCrAs presents a duality in that it acts ``normally" otherwise: the magnitude of the room temperature resistivity is typical of metallic conduction in a ternary transition metal compound; the low-temperature specific heat is linear~\cite{wu2009novel}; and the Wiedemann-Franz law is obeyed in the $T \rightarrow 0$ K limit~\cite{hope2021heat}.
    
    To reveal more information about the non-metallic resistivity and its potential connection to low-temperature non-Fermi liquid properties, optical conductivity measurements were done by Akrap et al.~\cite{akrap2014optical}. They found two distinct Drude components,
     one that is strong but broad, and the other weaker but narrower:
at room temperature, the plasma frequency and lifetime of the former
     are $\sim 3.5\times 10^{14}{\rm\ Hz}$ and $\sim 0.02$ picoseconds,
     while for the latter these values are
     $\sim 4.1 \times 10^{13}{\rm\ Hz}$ and $\sim 0.5$ picoseconds.
With decreasing temperature, the scattering rates follow
    the dc-resistivity quite well,
    showing that the rising resistivity is a result of
    anomalously increasing scattering, rather than loss of carriers due to gapping of the Fermi surface.
    
    One possible physical explanation that has been suggested, which is consistent with the optical conductivity measurements, is that FeCrAs is an extreme example of a so-called ``Hund's metal" system - systems which are characterized by the development of electron correlations through interorbital interactions~\cite{georges2013strong}. These interactions can compete with Mott-insulating states, resulting in large temperature ranges where electronic transport is incoherent~\cite{georges2013strong}. Within this regime, anomalous properties are predicted, for example: non-Fermi liquid scaling in transport properties~\cite{yin2012fractional}, suppression of Kondo temperatures~\cite{nevidomskyy2009kondo}, and the realization of a ``bad metal", where the Motte-Ioffe-Regel limit is violated~\cite{georges2013strong}. In FeCrAs, the product of the Fermi wavevector ($k_F$) and the mean-free-path ($l$) is estimated by combining DC resistivity measurements~\cite{wu2009novel} and band-structure calculations~\cite{akrap2014optical}, and is approximately 6. While this does not violate the Motte-Ioffe-Regel limit ($k_Fl \sim 1$), it shows that the scattering in FeCrAs is very strong, and thus some electron quasiparticles at the Fermi surface are not very well-defined.
    
    Another possible contributor towards anomalous scattering may arise from magnetic frustration~\cite{wang2016resistivity}. This frustration arises because the Cr magnetic moments sit upon a distorted Kagome sub-lattice. Although they order into a commensurate spin-density wave configuration below $T_N \sim 125$ K~\cite{wu2009novel}, due to frustration the magnetic energy scale implied by $T_N$ does not reflect the (much larger) exchange energies observed in inelastic neutron scattering measurements, which have observed magnetic fluctuations with an energy scale of around 900 K~\cite{plumb2018mean}. 
    
    Other studies have been done to distinguish between collective magnetic and local scattering: resistivity under hydrostatic pressure~\cite{tafti2013non}, muon-spin relaxation~\cite{huddart2019local}, and most recently, neutron diffraction~\cite{jin2019spin}. While a clear idea of the mechanisms behind the exotic behaviours of FeCrAs remains elusive, these experiments have shed some light on what may affect the non-metallic resistivity. In particular, resistivity measurements under hydrostatic pressure have shown that the non-metallic state is quite resilient, being relatively unaffected by pressures of up to 17 GPa, which are sufficient to suppress $T_N$. This suggests that even though magnetic order affects electronic scattering, as discussed below, the spin-density wave itself is not responsible for the non-metallic behaviour. 
    
    Here, we investigate the scattering mechanisms of FeCrAs through Hall effect measurements. We present measurements of the  Hall coefficient in single crystal FeCrAs, for current along the $c$-axis and within the hexagonal plane. Due to the high scattering rate in FeCrAs (and nearness to the Motte-Ioffe-Regel limit) our measurements are in the low-field limit, and are thus sensitive to the anisotropy of the scattering rate.  
    
    Extracting the low-field Hall coefficient from 2.6 K to 290 K, we find that the temperature dependence is anomalous. There is non-monotonic behaviour below $T_N$ in the form of two temperature-induced sign-reversals when the current is within the hexagonal plane. We interpret this as scattering from spin fluctuations that produce a momentum-dependent scattering rate below $T_N$. Howevever, this scattering seems to operate in parallel with some other, possibly Hund's mechanism, that operates over a much larger temperature range. 

\section{Experiment}
    Samples are from the same batch as in Ref.~\cite{wu2009novel} and two crystal orientations were studied: one where the magnetic field $B$ is within the $ab$-plane, with the excitation current along the $c$-axis, and vice-versa. The samples will henceforth be referred to by the direction of the excitation current: sample $c$ and sample $ab$ as in the order above. 
    
    An additional sample with the same orientation as sample $c$ was studied, and for that sample, the low-field Hall coefficient ($R_H$) qualitatively follows the same temperature dependence, albeit larger by a factor of approximately 1.5. It is likely that this difference in magnitude is due to uncertainty in the placement of soldered connections, uncertainties in sample thickness, and perhaps some minor sample dependence. We feel that the sample presented here is more reliable, as the Hall resistivity fluctuated less around its linear trend during magnetic field sweeps. Nevertheless, we focus on relative changes in $R_H$, which were the same in both samples, as opposed to its absolute magnitude.
    
    Measurements were carried out using a Quantum Design Physical Properties Measurement System via the AC transport option. 40 \si{\micro \metre} thick copper wires were carefully soldered onto the samples to minimize the size of the soldered connection. Since sample $c$ was larger in size (5.65 x 1.10 x 0.25 mm$^3$), it allowed the use of a balancing resistor, the so-called ``5-lead method", to minimize misalignment voltages from the Hall leads. The same lead setup could not be used by sample $ab$ because of its smaller size (1.05 x 0.56 x 0.19 mm$^3$), so a traditional 4-lead approach was used. Voltages were measured between the Hall leads under a magnetic field sweep from +9.0 to -9.0 T for sample $c$ and +7.0 to -7.0 T for sample $ab$ in 0.5 T steps, for each temperature. We note that we measured $R_H$ for sample $c$ for many more temperatures, which helped guide the choice of measurement temperatures for sample $ab$. 
    
    To extract the low-field Hall coefficient, $R_H$, the voltage ($V_m$) is measured for pairs of positive and negative magnetic fields, and then subtracted. This process algebraically removes even-in-field components in $V_m$ to obtain the Hall voltage, $V_H = [V_m(B) - V_m(-B)]/2$, where $B$ is the applied magnetic field. To relate $V_H$ to $R_H$, we have
    \begin{equation}
        \rho_{H} = \frac{V_H(|B|)t}{I} = R_H|B|,
        \label{eq:Hallcoefflinreg}
    \end{equation}
    where $\rho_H$ is the Hall resistivity, $I$ is the excitation current, and $t$ is sample thickness. We apply linear regression to the data and $R_H$ is returned as the best-fit slope. To ensure the correct sign for $V_H$, and therefore $R_H$, runs using the same procedure and data analysis were calibrated with a standard copper sample provided by Quantum Design. 

\section{Results}

    \begin{figure*}
        \includegraphics[angle=0,width=12.9 cm]{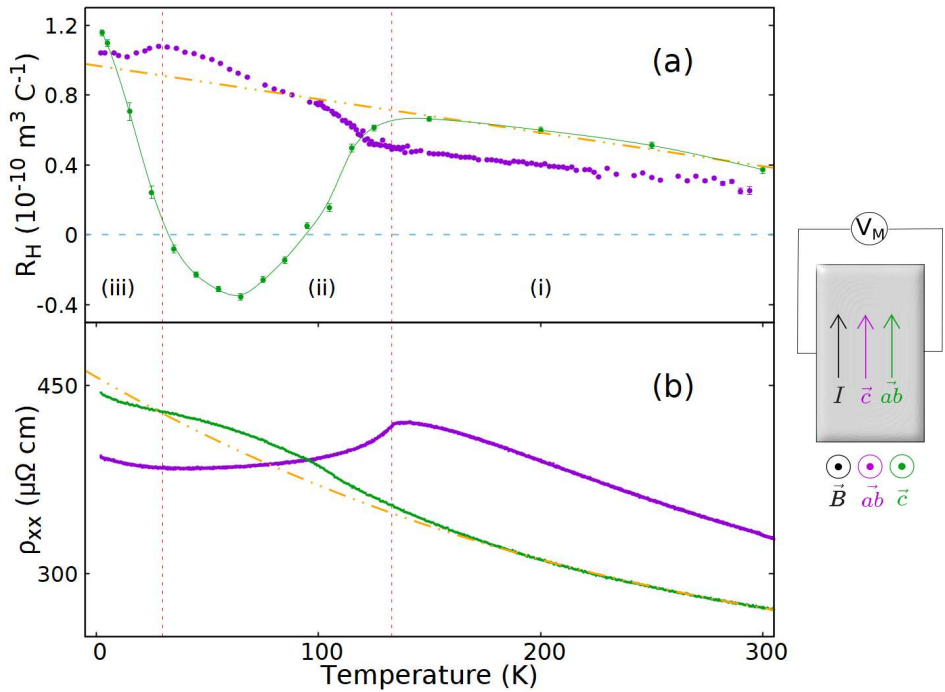}
        \caption{(Color online) The Hall coefficient and resistivity as a function of temperature are shown in plots (a) and (b) respectively, where the latter is taken from Wu et al.~\cite{wu2009novel}. Purple and green points correspond to data where the direction of the excitation current is along the $c$-axis and within the $ab$-plane, respectively; refer to the cartoon for a colored representation of the relationship between the current direction, magnetic field, and the crystal orientation. $V_M$ is the measured voltage for the Hall effect. The green curve in (a) is provided as a guide for the eye for sample $ab$ measurements, which are less dense in temperature. Dashed vertical lines section off the data into three regions, (i), (ii), and (iii), and a dashed horizontal line is provided to help identify zero-crossings. Orange double-dotted dashed lines are linear and cubic polynomial fits of high temperature transport within the $ab$-plane, in (a) and (b) respectively. Refer to text for further discussion.}
        \label{fig:hallcoeff}
    \end{figure*}

    The temperature dependence of $R_H$ is shown in Fig.~\ref{fig:hallcoeff}(a) for both samples. To aid discussion, $R_H$ for a particular orientation will be denoted by $R_H^{c}$ and $R_H^{ab}$ for sample $c$ and sample $ab$ respectively, and general reference to the Hall coefficient will remain $R_H$.
    
    We first separate the data into three temperature regions: (i), (ii), and (iii). Starting with $R_H^{c}$, we see that in region (i) it grows linearly with decreasing temperature. The boundary between regions (i) and (ii) coincides with the formation of static magnetic order at $T_N$. There is a sharp rise in $R_H^{c}$ starting at $T_N$ that extends down to approximately 90 K, where the behaviour again becomes roughly linear, but with a higher slope than in region (i). Region (iii) begins at around 30 K, where the temperature dependence begins to saturate, reaching a broad maximum, before eventually flattening to a roughly constant value at the lowest temperatures.
    
    Moving on to $R_H^{ab}$, the change in orientation has a drastic effect below $T_N$. In region (i), the behaviour of $R_H^{ab}$ is similar to that of $R_H^{c}$, where a quasi-linear rise is observed as the sample cools. $R_H^{ab}$ begins to fall a little above $T_N$, and at $T_N$, it changes rapidly, marking the start of region (ii). $R_H^{ab}$ decreases below $T_N$, becoming negative at around 100 K, reaching a minimum at about 65 K, and becoming positive again at around 35 K. This recovery towards positive values marks the end of region (ii), and correlates with the broad maximum of $R_H^{c}$. At the lowest temperatures, unlike in $R_H^{c}$, $R_H^{ab}$ does not saturate down to 2.6 K. Instead, $R_H^{ab}$ rapidly approaches values that are roughly consistent with an extrapolation of its (quasi-linear) high-temperature trend to the lowest temperatures, as shown by orange curve in Fig.~\ref{fig:hallcoeff}(a). 
    
    Due to the surprising temperature dependence of $R_H$ below $T_N$, we have also checked for the presence of an anomalous Hall effect (AHE). No deviation from linearity as a function of applied magnetic field was resolved in $V_H$ at any temperature. In the Supplementary Materials section, we show typical data at $T =$ 250, 125, 96, 36, and 2.6 K and conclude that within our experimental resolution, no AHE can be resolved, and thus the anomalous behaviour of $R_H$ is not due to a contribution from an AHE.

\section{Discussion}
    
    \subsection{Density wave gapping vs. scattering anisotropy}
    A strong temperature dependence of the low-field Hall effect is not uncommon even in elemental metals~\cite{cox1973temperature,saeger1968hall}, and it has been long understood as revealing the temperature dependence of anisotropic, or momentum-dependent, scattering~\cite{hurd1969hall,hurd1977anisotropic}. In the low-field limit, the quasiparticles on the Fermi surface scatter long before they can sample the entire cyclotron orbit. This makes the quasiparticles sensitive to momentum-dependent scattering rates, as well as to local Fermi surface curvature, as discussed below. 
    
    A rapid change in $R_H$ upon cooling, as seen in the transition between region (i) and (ii) in FeCrAs, typically represents one of two possible scenarios. The first, and perhaps less exotic, occurs when the Fermi surface becomes gapped due to the formation of density wave order; the second scenario is where there is a change to the momentum dependence of the scattering rate, i.e. the anisotropy of the scattering rate changes.
    
    Examples of density-wave-induced gapping can be seen in materials such as NbSe$_2$~\cite{lee1969low,naito1982electrical} and the superconducting cuprates~\cite{leboeuf2007electron}. In these materials, the response of $R_H$ to the development of the gap is quite strong and can produce sign-reversals in $R_H$, which are then followed by saturation towards a constant value at lower temperatures. This latter behaviour likely disqualifies Fermi surface reconstruction as an explanation for the anomalous behaviour in FeCrAs: in our measurement, $R_H^{ab}$ reverses sign twice. After the initial sign reversal around 100 K, further cooling causes $R_H^{ab}$ to recover towards positive values in region (iii). While it is possible that there is some gapping driven by the spin-density wave (SDW), the recovery of $R_H^{ab}$ towards positive values suggests that it cannot be the complete story. If Fermi surface reconstruction were the only mechanism for this anomalous behaviour, then exotic ideas such as re-entrant suppression of a Fermi surface gap would have to be explored, and the lack of significant changes to spectral density in optical conductivity measurements would have to be reconciled with such a picture.
    
    The other possibility, exemplified by Sr$_2$RuO$_4$~\cite{mackenzie1996hall}, is the rapid change of $R_H$ being driven by a momentum-dependence, i.e.\ anisotropy, of the scattering rate that is changing with temperature. In Sr$_2$RuO$_4$, as in FeCrAs, there are two temperature-induced sign-reversals in $R_H$: the sign changes from negative to positive upon cooling to 150 K, and remains positive until around 30 K, where it drops sharply towards negative values, eventually saturating at a constant negative value below 1 K. This scenario of temperature-dependent scattering rate anisotropy is the more plausible explanation for the non-monotonic temperature dependence of $R_H$ in FeCrAs. This interpretation does not require the Fermi surfaces to change drastically both at and well below $T_N$, and the introduction of scattering anisotropy is not farfetched, given the presence of frustrated itinerant magnetism in this material.

    While both FeCrAs and Sr$_2$RuO$_4$ exhibit anomalous temperature dependences in the form of two temperature-induced sign-reversals in $R_H$, the latter saturates at a negative value that is about ten times larger in magnitude than its high-temperature value. In contrast, $R_H^{ab}$ in FeCrAs does not saturate, and seems, as $T \rightarrow 0$~K, to recover towards a value that is consistent with an extrapolation from its high-temperature trend. This difference in behaviour implies that the sign changes seen in Sr$_2$RuO$_4$ reflect a permanent change in the scattering mechanism, upon cooling below a characteristic temperature, as suggested by Mackenzie et al.~\cite{mackenzie1996hall}, who applied the geometric Fermi surface analyses of Ong~\cite{ong1991geometric}. They suggest that there is a cross-over from an isotropic mean-free-path at low temperature where scattering is dominated by impurities, to an isotropic lifetime at high temperature. We note that $R_H(T)$ in Sr$_2$RuO$_4$ was more recently interpreted as a property of a Hund's metal by Zingl et al.~\cite{zingl2019hall}. In FeCrAs, that $R_H$ as $T \rightarrow 0$~K is similar to an extrapolated high-temperature value implies that the low-temperature scattering mechanisms in region (iii) are similar to the ones above $T_N$ in region (i). This suggests that the $T > T_N$ and perhaps the $T \rightarrow 0$ K regions reflect the persistent non-metallic scattering rate, as the anomalous $\rho(T)$ extends across a huge temperature range, with a temporary interruption below $T_N$. That is, inside region (ii), the picture is complicated by some additional scattering, which is anisotropic as demonstrated by our $R_H(T)$ data.
    
    \subsection{Correlation with zero-field resistivity}
    
    The complexity of the situation in FeCrAs is thus exacerbated by the possibility of two independent scattering processes - one that is driving the non-metallic temperature dependence in $\rho(T)$, and another that rides on top of it, which gives rise to what appears to be excess anisotropic scattering in region (ii).
    
    Evidence for two separate scattering rates in our measurements is first noticeable when we consider $\rho_a$, and see that the rising temperature dependence is weakly but noticeably interrupted as the system cools through $T_N$. This can be seen in Fig.~\ref{fig:hallcoeff}(b) comparing the orange curve, which represents an extrapolation of a cubic polynomial fit using $\rho_a$ data from 170 K to 600 K (full scale not shown) to the data (green line). Starting just above $T_N$, $\rho_a$ rises above the extrapolated curve, has a maximum deviation at a similar temperature ($\sim 60$~K) to where $R_H$ has its minimum, before returning towards the extrapolated curve as $T \rightarrow 0$~K. It seems that the excess scattering that appears around $T_N$ fades away again at the lowest temperatures. The similarity to the interrupted behaviour of $R_H^{ab}(T)$ is striking. Since the low-field Hall effect is sensitive to scattering rate anisotropy~\cite{hurd1972hall,gausepohl1996hall,ong1991geometric,ouisse2019modelling}, and the non-monotonic response seen in $R_H^{ab}$ correlates well with the post-SDW order rise in $\rho_a$, the two behaviours may have the same origin. This implies that the anomalous temperature dependence of $R_H^{ab}$ is a result of an additional, anisotropic, scattering mechanism below $T_N$ that leads to excess scattering seen in $\rho_a$.

    A likely physical mechanism for this excess scattering is momentum-dependent spin-fluctuations that begin to develop near $T_N$. In particular, we focus on the Cr spins that begin to form a SDW below $T_N$, becoming ordered in a 120$^{\circ}$ configuration in real space with an ordering vector of (1/3, 1/3, 0) in reciprocal space. In real space, this corresponds to a commensurate spin-density wave in which the ordered moments vary in magnitude from 0.6 to 2.2 $\mu_B$~\cite{wu2009novel} at 2.8 K. It seems plausible that on Cr sites where the ordered moment is low, the fluctuating moment is high, and vice-versa; in other words, the ordered static SDW is accompanied by spin fluctuations that are out-of-phase with it in real space, and thus momentum-dependent. We speculate that, in region (ii), these commensurate fluctuations are highly thermally-populated and give rise to the momentum-dependent scattering rate, creating ``hot-spots" of scattering on Fermi surfaces. A similar scenario has been suggested to explain anomalous transport properties in the iron-based superconductors~\cite{prelovvsek2009analysis,breitkreiz2013transport,koshelev2016magnetotransport,bristow2020anomalous}. Considering the negative $R_H$ in region (ii), this would require that hole-like regions of the Fermi surface scatter more strongly from these spin-fluctuations. One particular material of interest to note is the Hund's metal, FeSe~\cite{kostin2018imaging}, which was recently shown to display anomalous temperature dependence in its magnetotransport~\cite{bristow2020anomalous}. This robust behaviour was credited to scattering anisotropy within its nematic phase, and was suggested to be induced by spin fluctuations.
    
    Our picture gives a natural explanation for the revival of $R_H$ towards its high-temperature behaviour in region (iii). While it could simply be coincidence that $R_H(T \rightarrow 0~\text{K})$ is similar in magnitude to $R_H(T > T_N)$, for both $R_H^{ab}$ and $R_H^{c}$, it is interesting to surmise that this is because the magnetic environment at low temperatures has become relatively ``calm" again - spin fluctuations responsible for the anisotropic scattering rate are being thermally-frozen out as $T \rightarrow 0$~K.
    
    For completeness, we mention that below $T_N$, elastic neutron diffraction shows that ordered moments initially point along the $c$-axis, but upon further cooling the moments rotate towards the $ab$-plane. The moments settle into a fully in-plane orientation by 2.5 K, but non-zero $c$-axis components are present until at least 50 K~\cite{jin2019spin}. This spin reorientation coincides with the regime in region (ii), where $R_H^{ab}$ is negative, i.e. between $\sim$ 35 and 95 K. Thus, it is tempting to suggest that the two phenomena are linked, but there is no obvious mechanism.
    
    We finally note that the focus of the discussion in this section has been on $R_H^{ab}$ because of the subtler behaviour of $R_H^{c}$ and the difficulty of its analysis when taking $\rho_c$ into consideration. The overall scattering rate for $c$-axis current is suppressed below $T_N$ because the ordered Cr moments are parallel in the $c$-axis direction. The non-metallic contribution seen by $\rho_c$ is perhaps too well-hidden to compare to $R_H^{c}$ here. It is noteworthy, however, that the interpretation of excess scattering is consistent with measurements of $\rho_c$ under hydrostatic pressure, done by Tafti et al.~\cite{tafti2013non}. Applying pressure to FeCrAs was found to suppress both $T_N$ and its corresponding depression of $\rho_c$; this strongly suggests that the change in scattering seen below $T_N$ is directly tied to the SDW order, and that the non-metallic scattering rate is a separate phenomenon. 
    
    \subsection{Comparison with calculated Fermi Surface}
    
    \begin{figure}
    \centering
    \includegraphics[width = 8.6 cm]{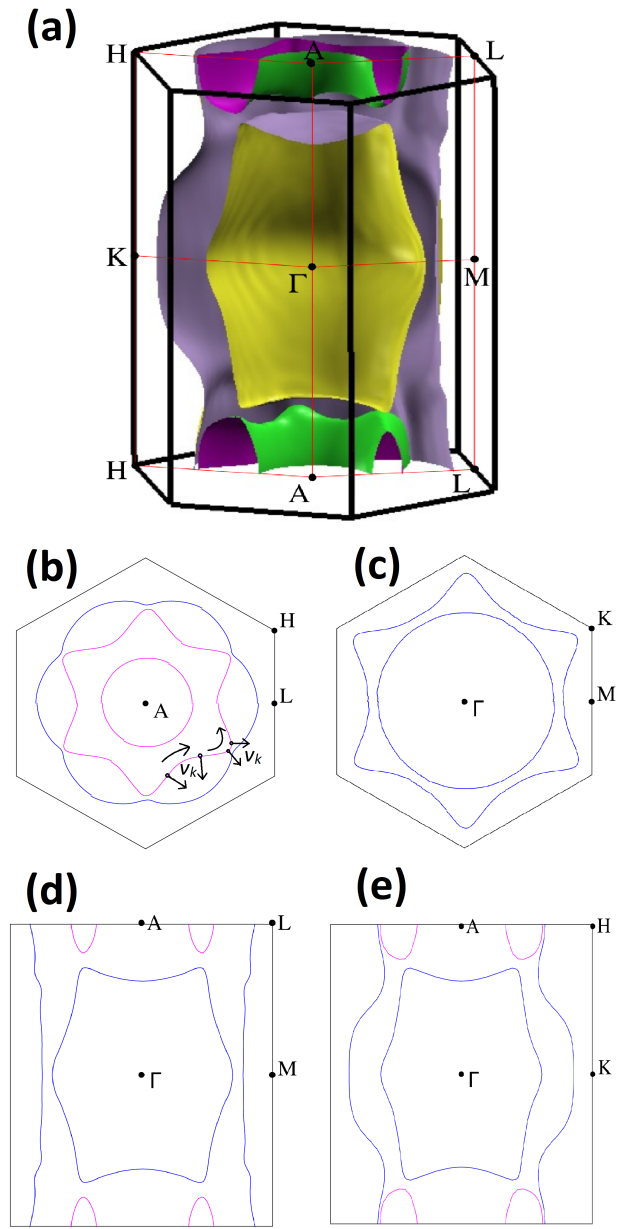}
    \caption{Cuts of the calculated Fermi surfaces of Akrap et al.~\cite{akrap2014optical}. There are three Fermi surfaces shown in (a): the outermost surface is a large distorted cylinder that is an electron surface; the closed surface centered on $\Gamma$ is a hole surface; and the torus near the top of the zone is an electron surface. (b) and (c) show two-dimensional cuts of the Fermi surfaces at the top and centre of the zone, respectively, within the hexagonal plane. Black arrows in (b) demonstrate time-evolution for the velocity of the state on the electron torus Fermi surface when the magnetic field is pointing out-of-the-page, demonstrating the ability of an electron surface to contribute hole-like behaviour in the low-field limit of the Hall effect (see text for discussion of how this applies to FeCrAs, or to Ong~\cite{ong1991geometric} and Ouisse et al.\cite{ouisse2015magnetotransport,ouisse2019modelling} for a more general discussion). (d) and (e) show two-dimensional cuts perpendicular to the hexagonal plane.}
    \label{fig:projFS}
    \end{figure}
    
    \begin{figure}
    \centering
    \includegraphics[width = 8.6 cm]{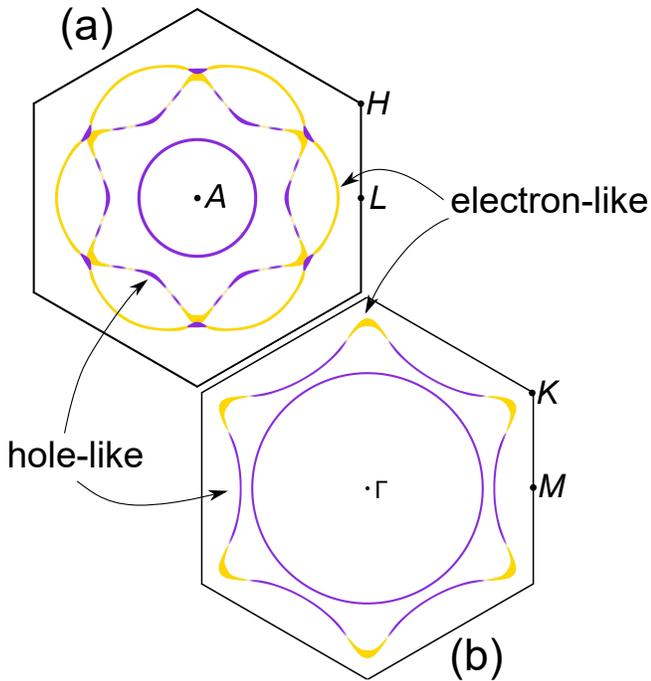}
    \caption{(Color online) The effects of curvature on the in-plane cuts of the Fermi surfaces. Violet and yellow regions denote electron-like and hole-like contributions to $R_H$, respectively, where thicker lines indicate a more intense curvature and thus a stronger contribution to $R_H$. The $A$-centred electron surface seen in (a) contains large regions that contribute a hole-like response to $R_H$. This is similar to the behaviour of the outer $\Gamma$-centred, electron surface seen in (b).}
    \label{fig:cutcurves}
    \end{figure}
    
    We now discuss in more detail the idea of anisotropic scattering in the low-field Hall effect, and how it relates to the calculated Fermi surface of FeCrAs. While we interpret our results as the development of ``hot-spots" below $T_N$, we show that the Fermi surface offers many possible locations for these hot-spots due to competing hole- and electron-like contributions on different Fermi surface sheets. These contributions cannot be resolved by only measuring $R_H$, but the difference in temperature dependence between $R_H^{ab}$ and $R_H^{c}$, as well as $\rho_a$ and $\rho_c$, suggest that these hot-spots are located on ``hole-like" regions where the Fermi velocity is predominantly in the $ab$-plane. 
   
   We begin by discussing Fermi surface curvature, and scattering using the geometric interpretation first introduced by Ong~\cite{ong1991geometric}, and more recently applied by Ouisse et al.~\cite{ouisse2015magnetotransport,ouisse2019modelling} to a hexagonal metal. As seen in Fig.~\ref{fig:projFS}, there are three Fermi surfaces to consider~\cite{akrap2014optical}: a pocket centred on $\Gamma$ that is a hole surface, a larger, open-ended cylinder that is an electron surface, and a star-shaped torus centered on $A$ that is an electron surface. The black arrows demonstrate the evolution of a state and the direction of its velocity on the torus when a magnetic field is applied out-of-the-page. Notice that the state circulates counter-clockwise as required by the Lorentz force, but the change to $\vec{v}_k$ in time does not necessarily follow that. In convex sections of the Fermi surface, the direction of $\vec{v}_k$ will evolve clockwise when the state itself moves counter-clockwise; this results in real-space movement that mimics the time-evolution of a state on a hole surface.
    
    In the high-field limit, this behaviour is inconsequential because the sign of $R_H$ is determined by whether the cyclotron orbit encloses filled or unfilled states. In the low-field limit, this is different - quasiparticle states scatter before they traverse much of the Fermi surface, and $R_H$ is sensitive to the momentum dependence of scattering and the details of the Fermi surface(s). As elegantly shown by Ong~\cite{ong1991geometric}, the relative contributions of electron- and hole-like regions depend on the local curvature and the mean free path $\vec{l}(\vec{k}) = \vec{v}_F(\vec{k})\tau(\vec{k})$, where $\vec{v}_F$ is the Fermi velocity. The only way for $R_H$ to change magnitude and sign is for the relative magnitude of $\vec{l}(\vec{k})$ on electron- and hole-like regions to change. If we assume $\vec{v}_F(\vec{k})$ to be independent of temperature, we conclude that in FeCrAs, the sign-reversal of $R_H$ in region (ii) is due to the enhancement of $\tau(\vec{k})$ on at least some hole-like regions of the Fermi surfaces. 
    
    We illustrate this in Fig.~\ref{fig:cutcurves} for the Fermi surfaces of FeCrAs, where (violet) hole-like and (yellow) electron-like contributions to $R_H$ are shown, and thicker lines indicate a larger curvature. The toroidal Fermi surface centred on $A$, and the outermost surface centred on $\Gamma$ are canonically ``electron surfaces" because they enclose filled-states. However, due to having convex curvature on much of the surface, they can contribute a hole-like response to $R_H$. In particular, in region (ii), we see that there is enhanced scattering in $\rho_{ab}(T)$, while $R_H^{ab}(T)$ also temporarily reverses in sign to become negative. This suggests that there are hot-spots of scattering developing, somewhere within the violet (hole-like) regions of the Fermi surfaces. We expect this interplay between curvature and enhanced anisotropic scattering to be driving both the sign-reversals and temperature dependence of $R_H$. 

\section{Conclusion}

    FeCrAs is a peculiar system that exhibits a ``non-metallic" resistivity, which suggests the existence of an anomalous scattering mechanism. Here, we measured the temperature dependence of the low-field Hall coefficient and observed that it is strongly temperature and orientation dependent below the onset of long-range magnetic order. We suggest that this is due to the presence of spin fluctuations that become momentum-dependent due to the static spin-density wave order, which provides a means of developing an anisotropic scattering rate. This anisotropy seems to exist for a limited temperature range below $T_N$, and vanishes as $T \rightarrow 0$ K. 
    
    We further interpreted our Hall measurements alongside previous optical conductivity and pressure studies below $T_N$, and argue that this is additional scattering that rides on top of the non-metallic scattering. We speculate that the two scattering rates have separate local and non-local origins. In the case of local scattering, we may be able to understand the non-metallic scattering, which acts over a huge temperature range, through significant Hund's coupling. However, non-local scattering, which acts below $T_N$, includes additional scattering off of spin fluctuations that change the momentum dependence of the scattering rate.
    
    Clearly, there are still unanswered questions regarding the unconventional scattering mechanism(s) in FeCrAs; what we revealed here is the presence of additional scattering anisotropy within the hexagonal plane that is separate from the previously observed non-metallic scattering rate. To better understand current data, theoretic treatments such as dynamical mean-field theory (DMFT) calculations would help us better understand contributions from Hund's coupling, as would experimental measurements of the $k$-dependent quasiparticle lifetime.
    
\section{Acknowledgements}
We are grateful to Y. J. Kim for assistance with the PPMS measurements, and to Y. B. Kim, and M. Fu for valuable discussions. We would like to acknowledge funding from NSERC (RGPIN-2019-06446).
    
\bibliographystyle{apsrev4-1}
\bibliography{bibby.bib}

\section{Supplemental Materials}
The linear relation in Eq.~\ref{eq:Hallcoefflinreg} is confirmed in the figure below for five representative temperatures where non-linearity might be suspected: far above the N\'eel temperature, near the N\'eel temperature, near the zero-crossings in $R^{ab}$, and at the lowest temperatures measured. We see that even-in-field components are successfully subtracted off, e.g. magnetoresistive components, and that a linear relationship holds with no resolvable contribution from the anomalous Hall effect.

    \begin{figure}[h!]
        \includegraphics[angle=-90, width=8.6cm]{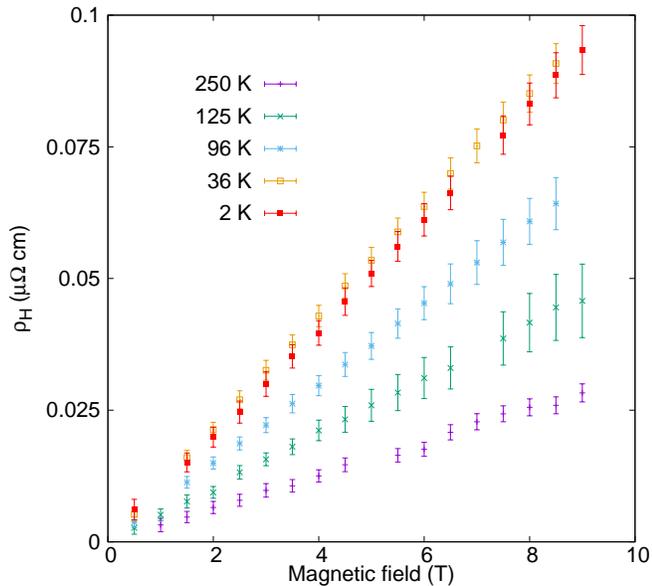}
        \caption{The Hall resistivity is shown to depend linearly on magnetic field. There are no resolvable deviations from linearity, and thus no evidence of an anomalous Hall effect. Data shown here are shown for five specific temperatures: 250, 125, 96, 36, and 2 K for sample $c$. We note that the linear-in-field dependence persists for all temperatures and for sample $ab$ as well.}
        \label{fig:halllinearity}
    \end{figure}

\newpage

\end{document}